\documentclass[12pt]{article}

\usepackage{times}
\usepackage{graphics}

\newenvironment{sciabstract}{%
\begin{quote} \bf}
{\end{quote}}

\title{ A Supernova Riddle}

\author {Douglas C. Leonard$^{1}$\\ 
\\ 
\normalsize {$^{1}$ The author is in the Department of Astronomy, San Diego
State University, 5500 Campanile}\\
\normalsize{Drive, San Diego, CA 92182-1221, USA.  E-mail:leonard@sciences.sdsu.edu}}

\date{}

\begin{document} 


\baselineskip20pt


\maketitle


\begin{sciabstract}

Analysis of the polarization of light from supernovae can reveal the shape and
distribution of matter ejected from exploding stars.  In this ``Perspectives''
commentary (published: 2007, {\it Science}, 315, 193), we review the young
field of Type Ia supernova spectropolarimetry and critically evaluate the
recent work of Wang et al. (2007, {\it Science}, 315, 212), in which a
suggestive trend is found in data from 17 Type Ia events.

\end{sciabstract}

% In setting up this template for *Science* papers, we've used both
% the \section* command and the \paragraph* command for topical
% divisions.  Which you use will of course depend on the type of paper
% you're writing.  Review Articles tend to have displayed headings, for
% which \section* is more appropriate; Research Articles, when they have
% formal topical divisions at all, tend to signal them with bold text
% that runs into the paragraph, for which \paragraph* is the right
% choice.  Either way, use the asterisk (*) modifier, as shown, to

\section*{}

Roughly once per second in the observable universe, a star explodes and
announces its death with an optical display that for weeks rivals the
brilliance of its parent galaxy.  These supernova events are classified into
several types, but among the most interesting are those called type Ia
supernovae (SNe Ia).  Astronomers' love affair with these beacons began in
earnest about a decade ago when two groups put them to work as distance
indicators and precisely mapped the expansion history of the universe well into
the regime that gravity was expected to have imprinted its decelerating
signature.  Instead, the data revealed a universe presently accelerating in its
expansion rate, a finding heralded by {\it Science} as the ``Scientific
Breakthrough of the Year'' in 1998 ({\it 1}), and one that has since survived
intense scrutiny and complementary experimental checks.  Yet for all the
fanfare and empirical success, it must be acknowledged that we are
fundamentally ignorant: We do not know how these stars explode.  On page 212 of
this issue, Wang {\it et al.} ({\it 2}) identify a suggestive trend in an
impressive set of SN~Ia data that may point the way towards a deeper
understanding of these enigmatic cosmic blasts.

Despite an embarrassing dearth of direct observational evidence, the first part
of the story of SNe Ia is largely considered settled.  Each future SN Ia begins
as a carbon-oxygen white dwarf -- the compact corpse of a low-mass star like
our Sun after its nuclear-burning life is over -- accreting matter through some
mechanism (mass flow from the envelope of a close companion star seems most
likely) until a critical central density is achieved and a thermonuclear
runaway is triggered.  There is general agreement that, once initiated, the
burning front progresses through the star for a time as a subsonic
deflagration.  But at this point in the story, harmony ends and pitched battles
begin, with some favoring an enduring deflagration front and others insisting
on a transition to a supersonic detonation.

The most recent ``delayed detonation'' models appear to better match observed
SNe~Ia: The events produced in these simulations are bright enough (a perennial
problem for deflagration models) and have the proper ejecta composition and
stratification ({\it 3}).  The mechanism that triggers the
deflagration-detonation transition remains a mystery, however, and so the pure
deflagration model still retains its share of adherents.  In any event, a
complete comparison of the observable distinctions predicted by the two
scenarios still awaits full, three-dimensional radiation transport simulations
carried out at high enough resolution to resolve physical processes at very
small scales.  Into this fray, Wang {\it et al.} now step, armed with an
upstart and potentially powerful observational tool: The ability to study the
{\it geometry} of the supernova ejecta by analyzing the polarization properties
of the light coming from the star shortly after explosion.

Are supernovae round?  Simple to pose, this question belies a menacing
observational challenge, given that all extragalactic supernovae remain
point-like in the night sky throughout the critical early phases of their
evolution.  Fortunately, geometric information is encoded in the polarization
properties of supernova light.  The essential idea is that photons become
polarized when they scatter off of free electrons, and hot, young supernova
atmospheres contain an abundance of free electrons.  Indeed, if we {\it could}
view such an atmosphere as an extended source, rather than as an unresolvable
point of light, we would expect to measure changes in both the direction and
strength of the polarization as a function of position in the atmosphere.  For
a spherical, unresolved source, the directional polarization components cancel
exactly and yield zero net polarization.  Any deviation from perfect symmetry
or roundness of the source in the plane of the sky, however, gives rise to a
net polarization (see the figure).

There are two basic causes of supernova polarization.  One is asphericity of
the electron-scattering atmosphere; because electron scattering is independent
of wavelength, it generally produces a uniform increase in the overall
polarization level across the spectrum.  In the other mechanism, asymmetry in
the distribution of material (``clumpy ejecta'') above the electron-scattering
photosphere unevenly screens the underlying light.  Unlike global asphericity,
this polarization mechanism is strongly dependent on wavelength, because only
those spectral regions corresponding to line transitions of the chemical
elements that make up optically thick clumps will be polarized.

From spectropolarimetry gathered on seven events, previous work in this young
field has found SNe~Ia to have low overall polarizations but occasionally
strong line polarization features ({\it 4 --- 7}).  The emerging picture is thus
one of a globally spherical photosphere with clumpy (or otherwise
asymmetrically distributed) ejecta overlying it.  How can such studies shed
light on the Type Ia flame-propagation mystery?  The latest models indicate
that pure deflagrations leave behind lumpier ejecta than delayed detonations do
({\it 3, 8}).

Spotting trends in SNe Ia data has a long tradition of bearing rich fruit.  In
1936, Walter Baade pointed out that the substantial homogeneity and
extraordinary brightness of these objects could make them powerful cosmological
tools.  By the early 1990s, however, it became clear that the dispersion in
peak intrinsic luminosity (by more than a factor of ten), complicated their use
as ``standard candles''.  The fix came in 1993, when Phillips ({\it 9})
quantified a trend first noticed by Pskovskii ({\it 10}) that intrinsically
bright SNe Ia rise and decline in brightness more slowly than dim ones do.
Various versions of the ``light curve-width'' relation have since provided the
edifice upon which the entire SN~Ia cosmology enterprise has been built, and
served as touchstones for theoretical models of the explosions.

It is just such a trend that Wang {\it et al.} now identify in
spectropolarimetry of 17 SNe~Ia: Bright events show systematically weaker line
polarization than dim ones do.  This trend is consistent with the idea that
different SNe~Ia make the transition from deflagration to detonation at
different times.  The sooner it happens, the brighter the supernova and the
more completely scoured the ejecta will be of the clumps left behind by the
deflagration front.  The agreement between model predictions and observations
strengthens the case for a detonation phase.

Will all debate now end on the subject?  It is doubtful.  Critics will point
out that the trend identified by Wang {\it et al}. specifically excludes all
spectroscopically ``peculiar'' SNe~Ia, which may comprise upwards of 30\% of
the total population ({\it 11}).  Fundamental advances often come from
consideration of the differences seen in a sample, rather than from the
similarities alone.  And some are likely to withhold any judgment until full
three-dimensional models capable of resolving the clumps and quantitatively
tracking the resulting polarization become available.  Simply put, too many
mysteries still surround SNe~Ia for anyone to grow complacent.  An important
clue appears to have been wrested from nature, but we are not ready to resolve
the riddle of SNe~Ia just yet.

\begin{quote}
{\bf References and Notes}

\begin{enumerate}
\item J. Glanz, {\it Science}, {\bf 282}, 2156 (1998).
\item L. Wang, D. Baade, F. Patat, {\it Science}, {\bf 315}, 212 (2007);
  published online 30 November 2006 (10.1126/science.1121656).
\item V. N. Gamezo, A. M. Khokhlov, E. S. Oran, {\it Astrophys. J.}, {\bf 632},
  337 (2005).
\item D. C. Leonard {\it et al.}, {\it Astrophys. J.}, {\bf 632}, 450 (2005).
\item L. Wang {\it et al.}, {\it Astrophys. J.}, {\bf 591}, 1110 (2003).
\item D. A. Howell, P. H{\" o}flich, L. Wang, J. C. Wheeler, {\it
  Astrophys. J.}, {\bf 556}, 302 (2001).
\item L. Wang {\it et al.}, {\it Astrophys. J.}, {\bf 653}, 490 (2006).
\item M. Reinecke, W. Hillebrandt, J. C. Niemeyer, {\it Astron. Astrophys.},
  {\bf 391}, 1167 (2002).
\item M. M. Phillips, {\it Astrophys. J.}, {\bf 413}, 105 (1993).
\item Y. P. Pskovskii, {\it Soviet Astron.}, {\bf 21}, 675 (1977).
\item W. Li {\it et al.}, {\it Astrophys. J.}, {\bf 546}, 734 (2001).
\end{enumerate}
\end{quote}
\newpage

% For your review copy (i.e., the file you initially send in for
% evaluation), you can use the {figure} environment and the
% \includegraphics command to stream your figures into the text, placing
% all figures at the end.  For the final, revised manuscript for
% acceptance and production, however, PostScript or other graphics
% should not be streamed into your compliled file.  Instead, set
% captions as simple paragraphs (with a \noindent tag), setting them
% off from the rest of the text with a \clearpage as shown  below, and
% submit figures as separate files according to the Art Department's
% instructions.

\begin{figure}[h!]

\begin{center}
 \scalebox{0.6}{
 \rotatebox{-90}{
	\includegraphics{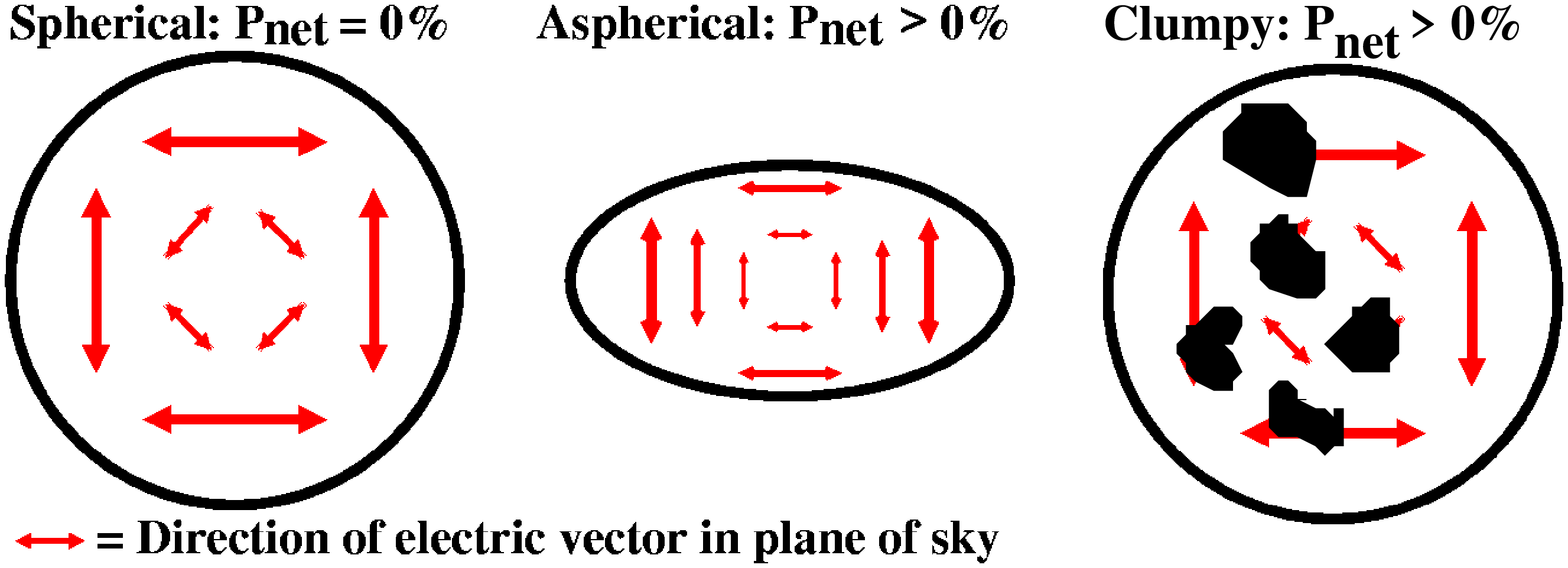}}}
\end{center}
\end{figure}

\noindent {\bf Producing supernova polarization.}  A spherical, unresolved
supernova atmosphere produces zero {\it net} polarization ({\bf left}), whereas
a non-spherical atmosphere does not ({\bf center}).  Clumps of material that
unevenly block the photosphere's light can also produce a net supernova
polarization ({\bf right}), and it is this mechanism that is thought to be
responsible for the majority of the observed polarization of SNe~Ia.

\end{document}